\documentstyle[aps,preprint]{revtex}
\begin{document}

\bibliographystyle{unsrt}

\title{ $\omega$ and $\rho$ Photoproduction with an Effective Quark Model Lagrangian}
\author{ Qiang Zhao $^1$, Zhenping Li$^{1}$, and C. Bennhold$^2$
\\
$^1$Department of Physics, Peking University, Beijing 100871, P.R.China
\\
$^2$Department of Physics, Center for Nuclear Studies,\\
The George Washington University, Washington D.C., 20052, USA}

\maketitle

\begin{abstract}
An unified approach for vector meson photoproduction is presented in the constituent 
quark model.  The s- and u-channel resonance contributions are generated
using an effective quark vector-meson Lagrangian.
In addition, taking into account $\pi^0$ and $\sigma$ t-channel
exchanges for diffractive production, 
the available total and differential 
cross section data for $\omega$, $\rho^0$, $\rho^+$, and $\rho^-$
photoproduction
can be well described with
the same quark model parameter set.
Our results clearly indicate that polarization observables
are essential to identify so-called ``missing" resonances.
\end{abstract}


\newpage
One of the main goals of the vector meson photoproduction experiments 
is to search for so-called ``missing" resonances, which have been
predicted by theory but have not been 
established experimentally\cite{missing,simon}.
One possible explanation for this long-standing puzzle is that these
states couple weakly to the $\pi N$ channel, which has provided most
information for $N^*$ states until now, but decay strongly into channels
like $\rho N$ and $\omega N$.
Encouraged by recent successful descriptions of 
pseudoscalar meson photoproduction\cite{pspaper}
we propose the parallel approach to vector meson 
photoproduction starting with the effective Lagrangian
\begin{equation}\label{1}
L_{eff}=-\bar {\psi}\gamma_\mu p^\mu \psi +\bar {\psi}\gamma_\mu e_q A^\mu \psi
+\bar {\psi} \left ( a \gamma_\mu + \frac {ib \sigma_{\mu\nu} q^\nu}{2m_q}\right
) \phi_m^\mu \psi +\dots 
\end{equation}
where $e_q$ ($m_q$) denote the quark charge (mass), $A^\mu$ the photon
field, and
where the quark field $\psi$ couples directly to the vector meson field
\begin{equation} \label{2} 
\phi_m =\left( \begin{array}{ccc}  
\frac{1}{\sqrt{2}}\rho^{0}+\frac{1}{\sqrt{2}}\omega & \rho^{+} & K^{*+}\\  
\rho^{-} & -\frac{1}{\sqrt{2}}\rho^{0}+\frac{1}{\sqrt{2}}\omega & K^{*0}\\  
K^{*-} & \overline{K}^{*0}  &\phi  
\end{array} \right )   
\end{equation}  
with momentum $q^\nu$. The coupling constants $a$ and $b$ in Eq. \ref{1}
allow for the two possible couplings of the quarks to the vector mesons;
they are free parameters to be determined by the data.  Unlike the large mass 
difference between the $\pi$ and $\eta$ in the peudoscalar case, the $\omega$
and $\rho$ states have nearly equal masses, thus isospin violations for the
$\omega$ and $\rho$ are relatively small. This encourages us to pursue an 
unified description of both $\omega$ and $\rho$ photoproduction with a
single set of parameters, where the vector mesons couple directly to the
quarks inside the baryon. 

We briefly outline our quark model approach to vector meson
photoproduction below;  a detailed derivation of the formalism 
is given in Ref. \cite{zq}.  Based
on the effective Lagrangian in Eq. \ref{1}; at tree-level
there are s-, u- and t- channel 
contributions, thus the matrix element for the meson photoproductions can 
be written as 
\begin{equation}\label{3}
{\cal M}_{fi}={\cal M}_t+{\cal M}_s+{\cal M}_{u}.
\end{equation} 

The derivation of the s- and u- channel contributions
uses methods similar to previous calculations of
pseudoscalar meson photoproduction\cite{pspaper}.
We separate the s-channel contributions ${\cal M}_s$ in 
Eq. 3 into two parts;
the s-channel resonances below 2 GeV and those above 2 GeV that
are treated as continuum contributions.
The electromagnetic transition amplitudes of s-channel baryon resonances and
their mesonic decays have been investigated extensively in the quark 
model\cite{cko,LY,simon,close} in terms of helicity and
the meson decay amplitudes.  These transition amplitudes for s-channel
resonances below 2 GeV have been translated into the standard helicity
amplitudes\cite{frank}  amplitudes in Ref.\cite{zq} in the harmonic
oscillator basis.
The framework of vector meson photoproduction in terms of the helicity amplitudes
has been thoroughly 
investigated\cite{frank}, and the various observables
can be easily evaluated in terms of these amplitudes.
The resonances above 2 GeV are treated as degenerate, since little
experimental information is available on those states.
Qualitatively, we find that the resonances with higher partial waves
have the largest contributions as the energy increases.  Thus, we
write the total contribution from all states belonging to the same 
harmonic oscillator shell in a compact form, 
using the mass and total 
width of the high spin states, such as $G_{17}(2190)$ for the $n=3$
harmonic oscillator shell.

The u-channel contributions ${\cal M}_u$ in Eq. \ref{1} include
the nucleon, the $\Delta$ resonance for 
$\rho$ production,  whose transition amplitudes
are treated separately, and all other excited states.  The excited states
are treated as degenerate in this framework, allowing their total 
contribution to be written in compact form.  This
is a good  approximation since contributions from u-channels 
resonances are not sensitive to their precise mass positions.   

The  t-channel exchange, ${\cal M}_t$,
is proportional to the charge of the outgoing mesons and
is needed to ensure gauge invariance of the total transition amplitude.
In addition to the t-channel exchanges from the 
effective Lagrangian in Eq. \ref{1}, additional t-channel exchanges are 
required for $\omega$ and $\rho^0$ prodcution in order to account for
the large diffractive behavior in the small scattering angle region. 
Using reasonable coupling constants 
Friman and Soyeur proposed\cite{fs} that such additional t-channel exchanges
can be described using $\pi^0$ exchange in the case of $\omega$ photoproduction
and dominantly $\sigma$ exchange in $\rho^0$ photoproduction.
Thus, we have also included the
$\pi^0$ and $\sigma$ exchanges for  $\omega$ and $\rho^0$ production,
respectively, but employed a form factor  at the corresponding vertices
that is given by the harmonic oscillator quark model 
 wavefunction.  This
leads to two additional parameters, $\alpha_{\pi^0}$ and $\alpha_{\sigma}$,
associated with the harmonic oscillator strength for the $\pi^0$ and $\sigma$
contributions. A detailed derivation of the $\pi^0$ exchange 
is given in Ref. \cite{zq1}.
 
We assume that the relative strengths and phases of each s-,
u- and t-channel term are determined by the quark model wavefunction
in the exact $SU(6)\otimes O(3)$ limit.
 The masses and decay widths
of the s-channel baryon resonances are obtained from the recent particle
data group\cite{pdg96}. The quark masses $m_q$ and the  parameter 
$\alpha$ for the harmonic oscillator wavefunctions in the quark
model are well determined in the quark
 model, they are
\begin{eqnarray}\label{4}
m_{u}=m_{d}=0.33 & \quad & \mbox{GeV} \nonumber \\ 
\alpha=410 & \quad & \mbox{MeV}.
\end{eqnarray} 
The coupling constants
for the $\pi^0$ and $\sigma$ exchanges are taken from Ref. \cite{fs}.  
This leaves only the coupling constants $a$ and $b$, and
the parameters $\alpha_{\pi}$ and $\alpha_{\sigma}$ to be determined 
by the data.

In Fig. 1, we compare total cross section data for 
$\gamma p\to \omega p$ and the three channels in $\gamma N\to \rho N$
with our calculations.
We did not perform a systematic fitting procedure due to the poor
quality of the data. Our study suggests that 
\begin{eqnarray}\label{5}
a &= & -1.7 \nonumber \\ 
b^\prime & = & b-a = 2.5 \nonumber \\
\alpha_{\pi} & = & 300 \quad \mbox{MeV} \nonumber \\
\alpha_{\sigma} & = & 250 \quad \mbox{MeV}
\end{eqnarray} 
leads to good overall agreement with the available data. 
Our results for the s- and u- channel
contributions alone are also shown for the reactions.
In general, the
contributions from the s- and u-channel resonances in $\omega$ and 
$\rho^0$ photoproduction account for only  20 to 40 percent of the total
cross section, demonstrating the dominance of diffractive scattering 
in these processes. Nevertheless, in the case of
$\omega$ photoproduction the quark model result
exhibits some resonance structure around 1.7 GeV photon lab energy
which comes from the $F_{15}(2000)$ state. A similar 
structure also appears in $\rho^0$ photoproduction, and additional 
contributions from the $F_{37}(1950)$, $F_{35}(1905)$, 
$P_{33}(1920)$ and $P_{31}(1910)$ resonances 
leads to a  broader structure. 
Clearly, the presence of diffractive scattering
complicates the extraction of 
the nucleon resonance contributions 
from the t-channel terms in the case of neutral
vector meson photoproduction.
Here, the photoproduction of charged vector mesons,
$\rho^-$ and $\rho^+$, presented in Fig. 1-c and 1-d,
become very important.  In these cases, the diffractive 
contributions are absent, and therefore,
resonance contributions dominate the
cross sections.  
Our numerical results for charged $\rho$ production are in good
agreement with the few  available data, even though the poor quality
of the data limit any conclusions that can be drawn.
Note that the cross section for
charged $\rho$ production is smaller by about a factor of three
compared to $\rho^0$ production.
Once the t-channel terms are added as described above
we also obtain a good description of the more numerous
$\omega$ and $\rho^0$ production data.  

The results for the differential cross sections for $\omega$ and $\rho$ production
are shown in Fig. 2. We find that the overall agreement with the available data for 
the differential cross sections is quite good as well.  
 As expected, the $\pi^0$ and $\sigma$ exchanges 
are responsible for the small-angle diffractive behavior, while the s- and 
u-channel resonances dominate the large momentum transfer region.
We point out that 
$\rho^-$ and $\rho^+$ production also shows some the diffractive behavior, 
although the size of the effect is smaller compared to $\omega$ and $\rho^0$
production.  This behavior can be explained by t-channel $\rho^-$ or $\rho^+$
exhanges, which are naturally generated 
 by the effective Lagrangian in Eq. \ref{1}.
The data in the reaction $\gamma n\to \rho^- p$ are in very good agreement
with the quark model predictions, indicating that the quark model
wave functions appear to provide the correct relative strengths and phases among 
the terms in the s-, u- and t-channels.

While the shapes and magnitudes of the differential cross sections are
well reproduced within our approach we find little sensitivity to
individual resonances.  For example,
in the energy region of $E_{\gamma}\sim 1.7$GeV, removing the
$F_{15}(2000)$ state - one of the ``missing" candidates -
changes the cross section very little, indicating
the differential cross section may not be the
 ideal experimental observable to study the structure
of the baryon resonances.
In contrast to the cross sections, the polarization observables
show a more dramatic dependence on the presence of the s-channel
resonances. To illustrate their effects
we show, as an example, the target polarization for the four channels
in $\omega$ and $\rho$ production with and without the contribution from
the $F_{15}(2000)$ resonance.  We do not expect the quark model in the 
$SU(6)\otimes O(3)$ limit to provide a good description of these observables.
However, it demonstrates the sensitivity of  these observables
to the presence of s-channel resonances.
Fig. 3 shows that the $F_{15}(2000)$ resonance has the
most dramatic impact on the $\omega$ channel while the effects
on the $\rho$ channels are smaller due to the contributions from the isospin
3/2 resonances, $F_{37}(1950)$, $F_{35}(1905)$, $P_{33}(1920)$ and 
$P_{31}(1910)$, which reduce the significance of the
$F_{15}(2000)$ state.  This shows that  polarization observables are essential
in analyzing the role of s-channel resonances.

In summary,  this investigation presents the first attempt to describe 
$\omega$ and $\rho$ meson photoproduction in a quark model
plus diffractive scattering framework.
It establishes the connection between the reaction
mechanism and the underlying quark structure of the baryons resonances. 
The crucial role played by the polarization observables in determining
the s-channel resonance properties is demonstrated.
Data on these observables, expected from TJNAF in the near future,
should therefore provide new insights into the
structure of the resonance $F_{15}(2000)$ as well as other ``missing" resonances.

One author (Z. Li) acknowledges the hospitality of the
Center for Nuclear Studies at The
George Washington University.  Discussions with F.J. Klein regarding
the data are also acknowledged.
This work was supported in part by the Chinese Education Commission and the 
US-DOE grant DE-FG02-95-ER40907.

\subsection*{Figure Caption}
\begin{enumerate}
\item The total cross section for (a): $\gamma p\to \omega p$, (b): $\gamma
p\to \rho^0 p$, (c): $\gamma n\to \rho^- p$, and (d): $\gamma p\to \rho^+
n$. The short-dashed line in (a) and (b) corresponds to the contributions
from the transition matrix elements generated with the effective
Lagrangian in Eq. 1, while the dashed line in (c) represents to cross
section for $t\le 1.1$ GeV$^2$. The data in (a) and (b) come from 
Ref. \cite{data}(triangle) and Ref.\cite{olddata}(square).
The data in (c) were taken with
the restriction  $t\le 1.1$ GeV$^2$ given by Ref.\cite{benz},
and the data in (d) come from Ref.\cite{barber}.

\item The differential cross section for (a): $\gamma p\to \omega p$ at
$E_\gamma =1.675$ GeV, (b): $\gamma p\to \rho^0 p$ at $E_\gamma =1.730$ 
GeV, (c): $\gamma n\to \rho^- p$ at 
$E_\gamma =1.850$ GeV, and (d): $\gamma p\to \rho^+n$ at
$E_{\gamma}=1.850$. The short-dashed line in (a) and (b) denotes the contributions
from the terms generated by the effective Lagrangian in Eq. 1. while 
the dashed line denotes the contributions from the diffractive processes.
The experimental data in (a) and (b) come from Ref. \cite{data}, and
in (c) come from Ref.\cite{benz}.

\item The target polarization for (a): $\gamma p\to \omega p$, (b): $\gamma
p\to \rho^0 p$, (c): $\gamma n\to \rho^- p$, and (d): $\gamma p\to \rho^+
n$ at $E_{\gamma}=1.7$GeV. The short-dashed lines show the result without the contribution
from the $F_{15}(2000)$.
\end{enumerate}


\begin{thebibliography}{99}
\bibitem{missing} N. Isgur and G. Karl, Phys. Letts. {\bf 72B}, 109(1977); Phys. Rev. {\bf D23}, 817(1981). 
\bibitem{pspaper} Zhenping Li, Ye Hongxing, and Lu Minghui, Phys. Rev. {\bf C56}
1099(1997).
\bibitem{zq} Q. Zhao, Z. P. Li, and C. Bennhold, nucl-th/9711061, submitted to Phys. Rev. {\bf C}.
\bibitem{fs} B. Friman and M. Soyeur, Nucl. Phys. {\bf A100}, 477(1996).
\bibitem{zq1} Q. Zhao, Z. P. Li, and C. Bennhold, ``Vector Meson Photoproduction
 with an Effective Lagrangian in the Quark Model II: $\omega$ Photoproduction",
 submitted to Phys. Rev. {\bf C}.
\bibitem{cko} L. A. Copley, G. Karl and E. Obryk, Nucl. Phys. {\bf B13},
303(1969); R. P. Feynman, M. Kislinger and F. Ravndal, Phys. Rev. {\bf
D3}, 2706(1971).
\bibitem{LY} Le Yaouanc {\it et al}, Hadron Transitions in the
Quark Model, (Gordon and Breach, New York, 1988); Phys. Rev. {\bf D8},
2223(1973); {\bf D9}, 1415(1974).
\bibitem{simon} R. Koniuk and N. Isgur, Phys. Rev. {\bf D21}, 1868(1980).
\bibitem{close}F. E. Close and Zhenping Li, Phys. Rev. {\bf D42},
2194(1990); Zhenping  Li and F.E. Close, {\it ibid.}, 2207(1990).
\bibitem{frank} M. Pichowsky, \c{C}. \c{S}avkl\i, F. Tabakin, Phys. Rev. 
{\bf C53}, 593 (1996).
\bibitem{pdg96} Particle Data Group, E. J. Weinberg, {\it et al}, Phys. Rev. {\bf D54}, 1(1996).
\bibitem{data} F. J. Klein, to appear on the Proceedings of the GW/TJNAF Workshop 
on $N^*$  Physics, (Washington, D.C.) 1997, F.J. Klein, Ph.D. thesis, University of Bonn (1996).
\bibitem{olddata} H. R. Crouch {\it et al}, Phys. Rev. {\bf 155}, 1468(1967);
Y. Eisenberg {\it et al}, Phys. Rev. {\bf D5}, 15(1972);
Y. Eisenberg {\it et al}, Phys. Rev. Lett. {\bf 22}, 669(1969);
D. P. Barber {\it et al}, Z. Phys. {\bf C26}, 343(1984);
J. Ballam {\it et al}, Phys. Rev. {\bf D7}, 3150(1973);
W. Struczinski {\it et al}, Nucl. Phys. {\bf B108}, 45(1976).
\bibitem{benz} P. Benz {\it et al}, Nucl. Phys. {\bf B79}, 10(1974).
\bibitem{barber} D. P. Barber {\it et al}, Z. Phys. {\bf C2}, 1(1979).
\end{thebibliography}
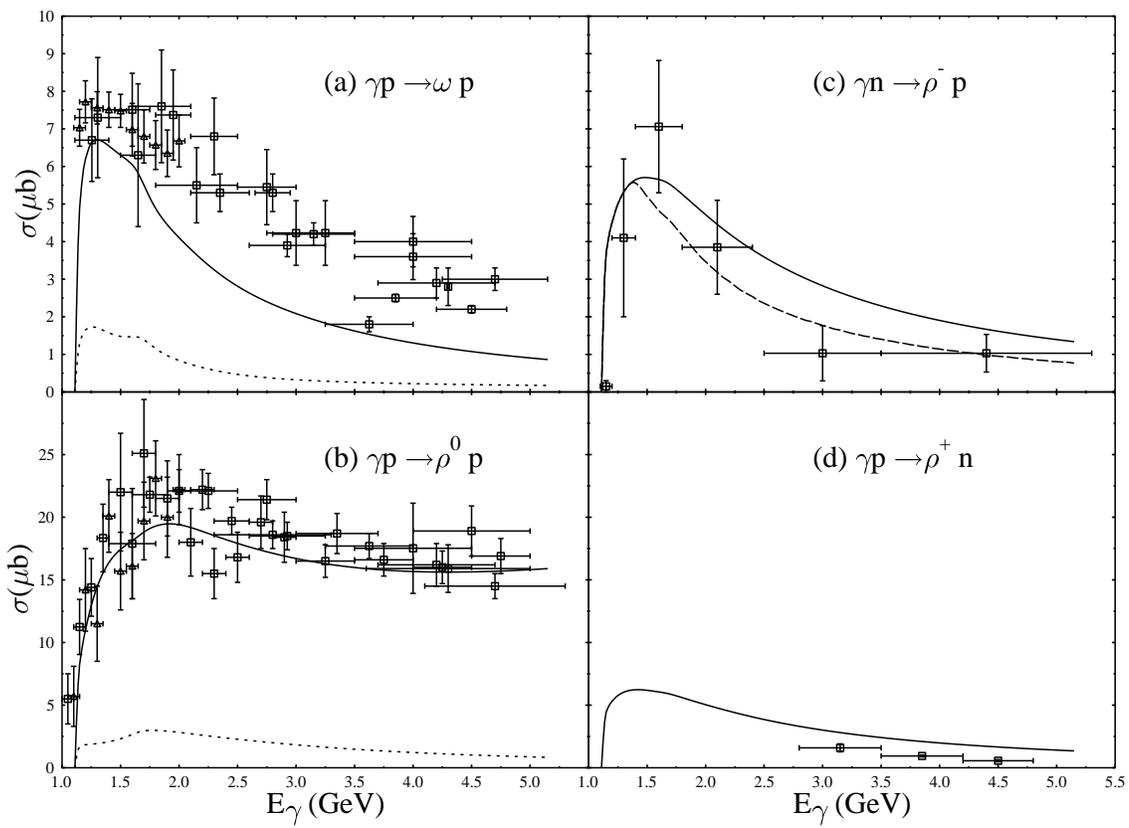
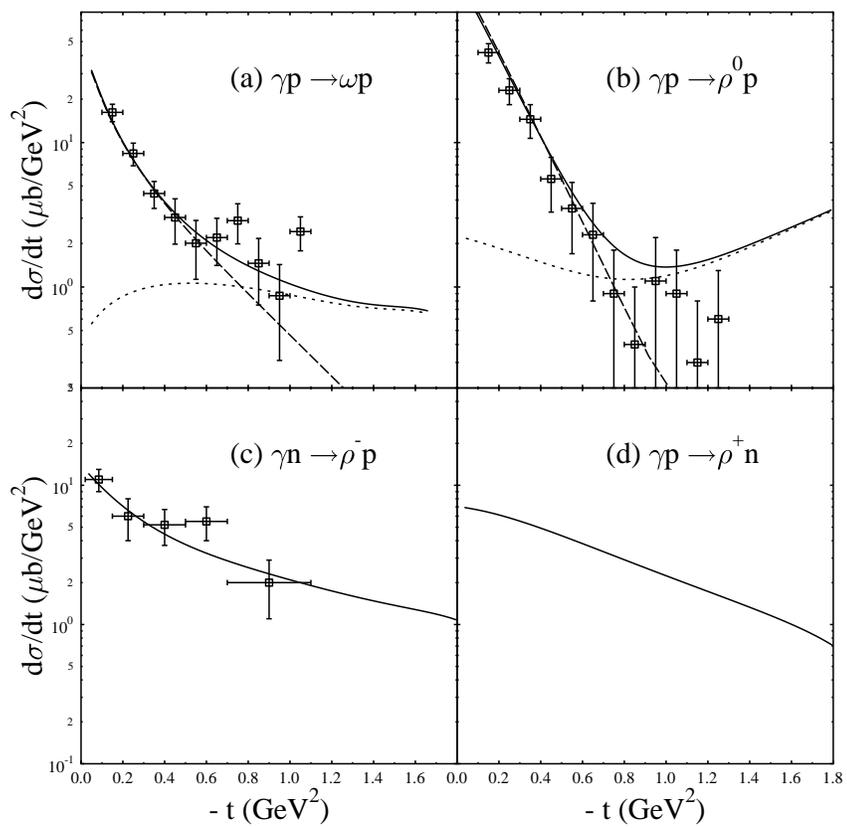
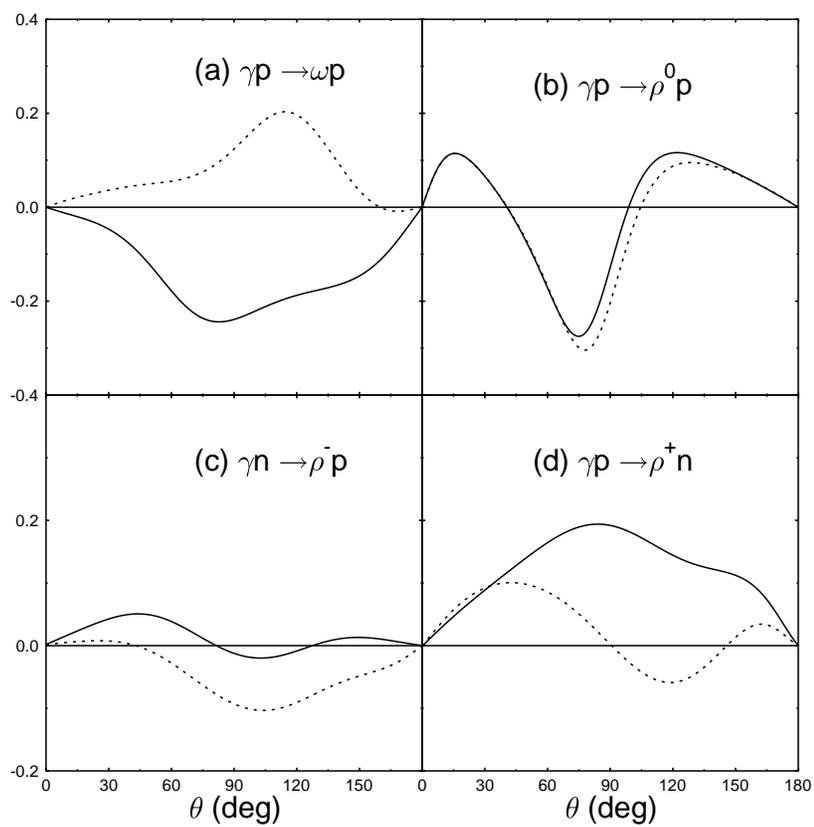
\end{document}